\documentclass[twocolumn]{aastex63}
\usepackage{graphicx}
\usepackage{dcolumn}
\usepackage{bm}
\usepackage{amsmath}

\begin{document}

\title{High-energy Neutrinos from Stellar Explosions in Active Galactic Nuclei Accretion Disks}

\author[0000-0002-9195-4904]{Jin-Ping Zhu}
\affil{Department of Astronomy, School of Physics, Peking University, Beijing 100871, China; \url{zhujp@pku.edu.cn}}

\author[0000-0003-4976-4098]{Kai Wang}
\affiliation{Department of Astronomy, School of Physics, Huazhong University of Science and Technology, Wuhan 430074, China; \url{kaiwang@hust.edu.cn}}

\author[0000-0002-9725-2524]{Bing Zhang}
\affiliation{Department of Physics and Astronomy, University of Nevada, Las Vegas, NV 89154, USA; \url{zhang@physics.unlv.edu}}

\begin{abstract}

Some catastrophic stellar explosions, such as supernovae (SNe), compact binary coalescences, and micro-tidal disruption events, are believed to be embedded in the accretion disks of active galactic nuclei (AGN). We show high-energy neutrinos can be produced efficiently through $pp$-interactions between shock-accelerated cosmic rays and AGN disk materials shortly after the explosion ejecta shock breaks out of the disk. AGN stellar explosions are ideal targets for joint neutrino and electromagnetic (EM) multimessenger observations. Future EM follow-up observations of neutrino bursts can help us search for yet-discovered AGN stellar explosions. We suggest that AGN stellar explosions could potentially be important astrophysical neutrino sources. The contribution from AGN stellar explosions to the observed diffuse neutrino background depends on the uncertain local event rate densities of these events in AGN disks. By considering {thermonuclear SNe, core-collaspe SNe, gamma-ray burst associated SNe, kilonovae, and choked GRBs} in AGN disks with known theoretical local event rate densities, we show that these events may contribute to $\lesssim10\%$ of the observed diffuse neutrino background.

\end{abstract}

\keywords{Cosmological neutrinos(338); High energy astrophysics (739); Active galactic nuclei(16); Supernovae(1668) }

\section{Introduction}
The accretion disks of active galactic nuclei (AGN) are believed to be the hosts for some massive stars and stellar remnants including white dwarfs (WDs), neutron stars (NSs), and black holes (BHs). One potential formation channel for such AGN stars and stellar remnants is the capture from the nuclear star clusters around the AGNs \citep[e.g.,][]{syer1991,artymowicz1993,fabj2020}. Furthermore, some AGN stars can be formed {\em in situ} in the self-gravitating region of the disk \citep[e.g.,][]{collin1999,goodman2003,goodman2004,wang2011,wang2012,dittmann2020}. These AGN stars will end up with supernovae (SNe) to pollute the disk with heavy elements, which can offer a possible explanation for the observational features of high-metallicity environment in AGN disks \citep[e.g.,][]{hamann1999,warner2003}, and hence leave behind some stellar remnants inside the disks. AGN disks provide a natural environment for embedded stars and compact objects to grow, to accrete materials, and to migrate within it \citep[e.g.,][]{bellovary2016,cantiello2021,dittmann2021,jermyn2021,wang2021a,kaaz2021,kimura2021,peng2021,pan2021}. There could be abundant stars and compact objects gathering in the inner part of the AGN disks, so that some stellar explosions can frequently occur there. 

Very recently, the field of stellar explosions in AGN accretion disks is exploding. The event rates and possible observable signatures for several kinds of stellar explosions occurring in AGN disks, such as SNe \citep{zhu2021c,grishin2021,moranchelbasurto2021}, gamma-ray bursts \citep[GRBs;][]{cheng1999,perna2021a,zhu2021a,zhu2021b}, NS mergers \citep{mckernan2020,zhu2021a}, binary BH (BBH) mergers \citep[e.g.,][]{mckernan2012,bartos2017,yang2019,yang2020,mckernan2019,graham2020,tagawa2020,tagawa2021,wang2021b}, Bondi explosions \citep{wang2021a}, micro-tidal disruption events \citep[mTDEs;][]{yang2021} and accretion-induced collapses of WDs \citep{zhu2021c} and NSs \citep{perna2021b}, have been investigated in detail. Different from the classical low-density environments in which these stellar explosions occur, the AGN environment allows the ejecta launched from AGN stellar explosions to interact with the AGN gaseous disk to drive an energetic shock. Such a shock can finally break out from the disk surface and power observable luminous electromagnetic (EM) signals \citep{zhu2021a,zhu2021c,grishin2021}. Neutrino emissions from the interaction between SN ejecta and the dense circumstellar medium have been studied intensively within the context of type II SNe \citep[SNe II, e.g.,][]{waxman2001,katz2011,murase2011,murase2018,murase2019,li2019,wang2019}. In principle, interaction of the ejecta from AGN stellar explosions with the AGN disk atmosphere can produce high-energy neutrino emission similar to those of SNe II. This motivates us to take the first step to study these new potential neutrino sources in this {\em Letter}.


\section{Model} \label{sec:model}
\subsection{Disk Structure}
The accretion disk model by \cite{sirko2003} gives a good description for the radial structure in the inner disk region, up to $\lesssim10^5\,r_{\rm S}$, where $r_{\rm S} \equiv 2GM_\bullet /c^2$ is the Schwarzschild radius of the central supermassive BH (SMBH), $G$ is the gravitational constant, $M_\bullet$ is the SMBH mass, and $c$ is speed of light. The disk can be supported by the orbital energy of the gas in the inner parts of the disk i.e., $r\lesssim 10^3\,r_{\rm S}$. The mid-plane radial density, always valid for the radial region of $10^{3}\lesssim r /r_{\rm S} \lesssim10^{5}$, can be expressed as
\begin{equation}
\label{Eq:RadialDensityProfile}
    \rho_0(r) = \frac{\Omega^2}{2\pi GQ} \approx 1.24\times10^{-9}\,{\rm g}\,{\rm cm}^{-3}\,M_{\bullet,8}^{-2}\left(\frac{r}{10^3\,r_{\rm S}}\right)^{-3},
\end{equation}
where $\Omega = (GM/r^3)^{1/2}$ and the Toomre parameter $Q = Q_{\rm min} \approx 1$ which assumes that the disk is heated by the release of orbital energy and auxiliary input energy due to the feedback of star formation. Hereafter, the convention $P_x = P/10^x$ is adopted in cgs units. At the same radial region, the disk scale height compared to the radial size of the disk is
\begin{equation}
\label{Eq:AspectRatio}
    \frac{H}{r} \approx 8\times 10^{-3} \left(\frac{r}{10^3\,r_{\rm S}}\right)^{1/2}.
\end{equation}
In the radial distance range of our interest, one can almost use Equation (\ref{Eq:RadialDensityProfile}) and (\ref{Eq:AspectRatio}) to calculate the mid-plane radial density and disk scale, respectively. 

We adopt a gas-dominated disk, and the vertical density profile is given by a Gaussian density profile \citep[e.g.,][]{netzer2013}
\begin{equation}
    \rho_{\rm d}(r , h) = \rho_0(r)\exp( -{h^2}/{2H^2} ),
\end{equation}
where $h$ is the vertical height to the mid-plane. 

\subsection{Shock Dynamics}
For an AGN stellar explosion with energy $E_0$ and ejecta mass $M_{\rm ej}$, a forward shock and a reverse shock are formed as the ejecta crashes into the disk atmosphere. We only consider the neutrino emission from the forward shock, since the contribution of the reverse shock are usually much weaker \citep{murase2011}. The shock will finally break out from the disk surface so that we focus on the vertical height $h$ above the mid-plane of the disk. By assuming that all events are mid-plane explosions, the shock velocity can be described as \citep{matzner1999}
\begin{equation}
    v_{\rm s}(h) \approx \left( \frac{E_0}{M_{\rm ej} + M(h)} \right)^{1/2}\left( \frac{\rho_{\rm d}(h)}{\rho_0} \right)^{-\mu},
\end{equation}
where $\mu\approx0.19$ and $M(h) \approx 4\pi\int_0^{h}\rho(h)h^2dh$ is the swept mass. The kinetic luminosity of the shock at $h$ is given by
\begin{equation}
    L_{\rm s} = 2\pi\rho_{\rm d}v_{\rm s}^3h^2.
\end{equation}
One can also calculate the time after an explosion when the shock moves to a vertical height of $h$ 
\begin{equation}
    t(h) \approx\int^{h}_0\frac{dh}{v_{\rm s}(h)}.
\end{equation}

\section{Neutrino Production} \label{sec:NeutrinoProduction}
The shock before breaking out may be radiation-dominated so that particle acceleration is prohibited \citep{waxman2001,murase2011,katz2011}. When photons start to escape and hence the shock is expected to become collisionless, particle acceleration and neutrino production can occur. The shock breakout takes place when $c / v_{\rm s}(h_{\rm bo}) \approx \tau(h_{\rm bo}) \approx \int_{h_{\rm bo}}^{+\infty}\kappa\rho(h)dh$, where $h_{\rm bo}$ is the breakout vertical height and we adopt a constant electron scattering opacity of solar composition, i.e., $\kappa \approx 0.34\,{\rm cm}^2\,{\rm g}^{-1}$, hereafter. The breakout time can be expressed as $t_{\rm bo} = t(h_{\rm bo})$.

After the shock breakout, protons could be accelerated to high energy via the Fermi acceleration mechanism with a power-law energy spectrum, $dn_p/d\epsilon_p \propto \epsilon_p^{-q}$ with $q\approx 2$ \citep{blandford1987,malkov2001}. The accreleration timescale is given by $t_{\rm acc} = \eta\epsilon_p/eBc$, where $\eta\sim20c^2/3v_{\rm s}^2$ is for Bohm limit, $e$ is the electron charge, and $B = \sqrt{4\pi\varepsilon_B\rho_{\rm d}v_{\rm s}^2}$ is the magnetic field strength in the shocked AGN disk material with $\varepsilon_B = 0.01$ being the typical value for the magnetic field energy fraction. 

The maximum proton energy ($\epsilon_{p,{\rm max}}$) depends on the comparison between the acceleration timescale and the cooling timescales. A high-energy proton mainly loses its energy through adiabatic loss and the inelastic hadronuclear reaction ($pp$). When the maximum energy of the accelerated protons is limited by $pp$ reaction, by equating $t_{\rm acc} = t_{pp} \approx 1 / [(\rho_{\rm s}/m_p)\kappa_{pp}\sigma_{pp}c]$, where $\rho_{\rm s} = 4\rho_{\rm d}$ is the density of the shocked disk material, $\kappa_{pp} \approx 0.5$ is the $pp$ inelasticity, and $\sigma_{pp} \approx 5\times10^{-26}\,{\rm cm}^{-2}$ is the $pp$ cross section \citep{particle2004}, we have the maximum energy
\begin{equation}
\begin{split}
     \epsilon_{p,{\rm max}}^{pp} &\approx 94\,{\rm TeV}\,\rho_{{\rm d},-11}^{-1/2}v_{{\rm s},9}^{3}, \\
     &\approx9.4\times10^3\,{\rm TeV}\,\rho_{{\rm d},-15}^{-1/2}v_{{\rm s},9}^{3}.
\end{split}
\end{equation}
If the acceleration timescale of proton is limited by adiabatic cooling timescale of the shock, i.e., $t_{\rm acc} = t_{\rm ad} \approx h / v_{\rm s}$, one obtains
\begin{equation}
\begin{split}
    \epsilon_{p,{\rm max}}^{\rm ad} &\approx 1.7\times10^5\,{\rm TeV}\,\rho_{{\rm d},-11}^{1/2}v_{\rm s,9}^2h_{14}, \\
    &\approx1.7\times10^3\,{\rm TeV}\,\rho_{{\rm d},-15}^{1/2}v_{\rm s,9}^2h_{14}.
\end{split}
\end{equation}
The efficiency of $pp$ reaction can be estimated as \citep[e.g.,][]{razzaque2004,murase2008}
\begin{equation}
    \begin{split}
        f_{pp} \approx t_{\rm ad}/t_{pp} &\approx 1.8\times 10^3\,\rho_{\rm d,-11}v_{\rm s,9}^{-1}h_{14},\\
        & \approx 0.18\,\rho_{\rm d,-15}v_{\rm s,9}^{-1}h_{14}.
    \end{split}
\end{equation}
Because the density of the disk atmosphere after shock breakout is so large, it is expected that the $pp$ reaction is efficient\footnote{We note that since the target photons from ejecta-disk material interactions and the disk photons at the radial locations of our interested have energies of $E_\gamma\lesssim100\,{\rm eV}$, the threshold proton energy for photomeson production reaction ($p\gamma$) would be $\sim m_\pi m_pc^4/E_\gamma \gtrsim 1\,{\rm PeV}>\epsilon_{p,{\rm max}}^{pp}$ when the shock breaks out of the disk. This implies that $p\gamma$ is not very relevant in our cases. Furthermore, since the kinetic energy and breakout velocity of the shock driven by GRB-SNe are relatively large, in some cases $p\gamma$ interactions may dominate over $pp$ interactions. However, both interactions have similar neutrino production rates so that the effect of different hadronic processes on our conclusions can be ignored.}. Since the disk density then decreases rapidly and hence the shock would accelerate due to the Sakurai law \citep{sarkurai1960}, the adiabatic loss may become more important while the shock kinetic luminosity and the efficiency of $pp$ reaction drop sharply. We adopt a type Ia SN (SN Ia) with typical energy $E_0 = 10^{51}\,{\rm erg}$ and ejecta mass $M_{\rm ej} = 1.3\,M_\odot$ occurring at $r=10^3\,r_{\rm S}$ around a SMBH of mass $M_\bullet = 10^7\,M_\odot$ as the fiducial model hereafter. As an example, when the shock breaks out, one can calculate the vertical height $h_{\rm bo} \approx 4.2H \approx 1.0\times10^{14}\,{\rm cm}$, the density $\rho_{\rm bo} = \rho_{\rm d}(h_{\rm bo}) \approx 1.5\times10^{-11}\,{\rm g}\,{\rm cm}^{-3}$, the shock velocity $v_{\rm bo} = v_{\rm s}(h_{\rm bo}) \approx 1.1\times10^9\,{\rm cm}\,{\rm s}^{-1}$, and the breakout time $t_{\rm bo} \approx 2.2\times10^5\,{\rm s}$. The density would decease from $\rho_{\rm d} \sim 10^{-11}\,{\rm g}\,{\rm cm}^{-3}$ at $h = h_{\rm bo} \sim 4.2H$ to $\rho_{\rm d}\sim 10^{-15}\,{\rm g}\,{\rm cm}^{-3}$ at $h \sim 6H$ for our fiducial model. The shock kinetic luminosity (the efficiency of $pp$ reaction) would decay from $L_{\rm s,bo}\sim10^{45}\,{\rm erg}\,{\rm s}^{-1}$ ($f_{pp}\sim10^3$) to $L_{\rm s} \sim10^{42}\,{\rm erg}\,{\rm s}^{-1}$ ($f_{pp}\sim0.1$) after $\Delta t\sim1.8H/v_{\rm bo}\sim4\times10^4\,{\rm s}$ of the shock breakout. A high-energy proton would produce a pion by $pp$ reaction, then the charged pion decays leading to neutrino production. One can predict that the neutrino luminosity would have a sharp decay similar to that of $pp$ reaction.

\begin{figure}
    \centering
    \includegraphics[width = 0.99\linewidth , trim = 65 30 95 60, clip]{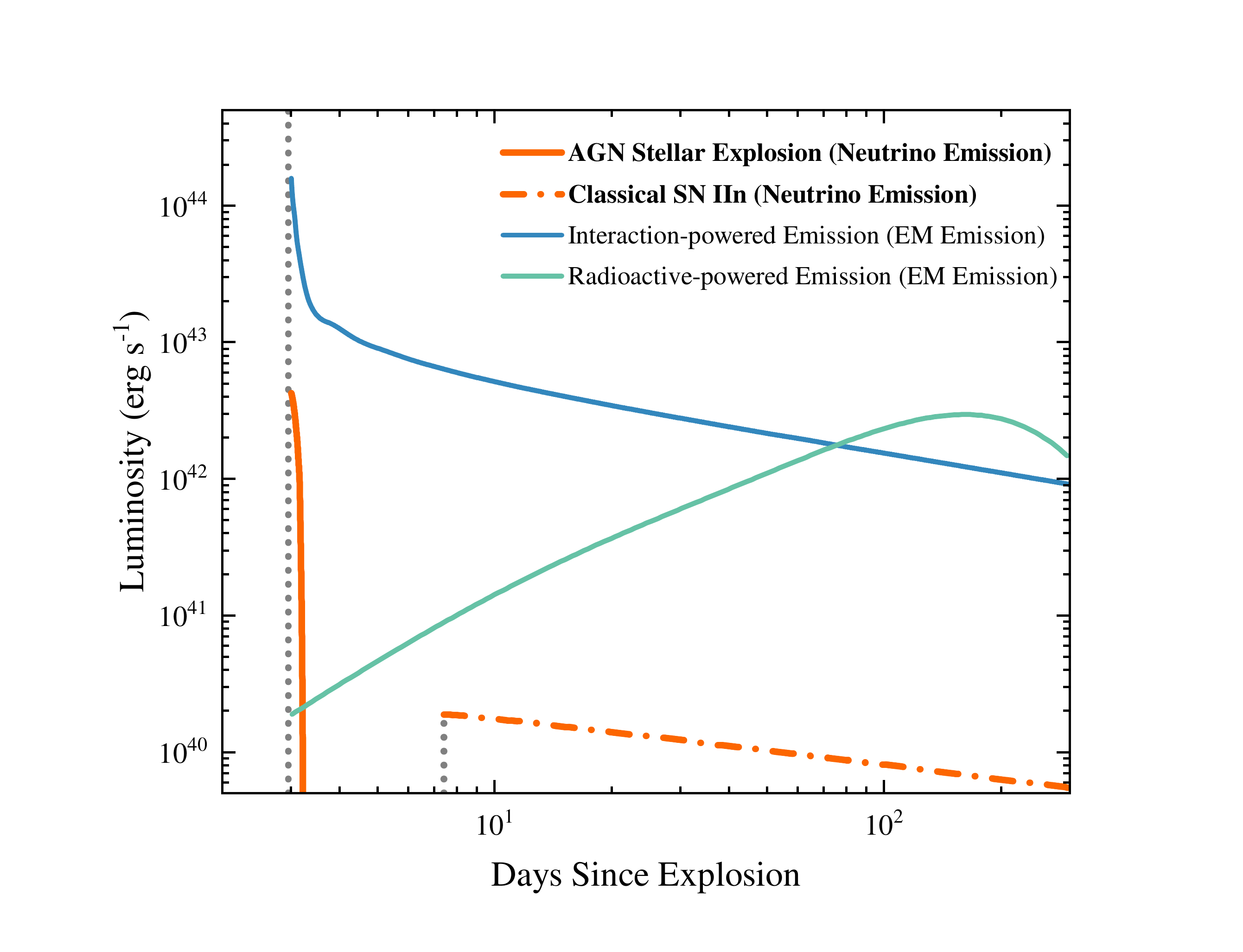}
    \caption{Lightcurves of neutrino emission (orange solid line; at $\epsilon_\nu = 1\,{\rm TeV}$) and EM emission for a SN Ia occurring at $r = 10^3\,r_{\rm S}$ around an SMBH of mass $M_\bullet = 10^7\,M_\odot$. The gray dotted lines represent the timescale of the shock breakout. The blue and green solid lines show the EM emission of the ejecta due to its interaction with the AGN disk materials \citep{grishin2021} and SN Ia emission powered by radioactive decay \citep{zhu2021c}, respectively. {For comparsion, we also show} the neutrino lightcurve for a classical SN IIn \citep{murase2018} (orange dash-dotted line). }
    \label{fig:MultiMessenger}
\end{figure}

We calculate the neutrino luminosity (for the sum of all flavors),
\begin{equation}
    \epsilon_\nu L_\nu \approx \frac{3K}{4(1+K)}\frac{\min(1 , f_{pp})\varepsilon_{\rm cr}L_{\rm s}}{\ln(\epsilon_{p,{\rm max}} / \epsilon_{p,{\rm min}})}, 
\end{equation}
where $\epsilon_\nu\simeq0.05\epsilon_p$, $K = 2$ denotes the average ratio of charged to neutral pion for $pp$ reaction, and $\varepsilon_{\rm cr} \approx 0.1$ is the energy fraction carried by cosmic rays \citep{caprioli2014}. We show the lightcurve of the neutrino emission at $\epsilon_\nu = 1\,{\rm TeV}$ for our fiducial model in Figure \ref{fig:MultiMessenger}. The duration of neutrino emission for an AGN SN Ia is much shorter than that of a classical SN IIn \citep{murase2018}, because AGN disks have a much sharper decaying density profile\footnote{There are existing disk winds \citep[e.g.,][]{proga2000,proga2004} and broad-line region \citep[e.g.,][]{moriya2017} clouds surrounding the accretion disks. The presence of disk winds and gaseous clouds would result in a relatively slower decaying density profile outside the disk, which can enhance the neutrino luminosity.} compared with that of the circumstellar medium around SNe IIn ($\rho\propto r^{-2}$). However, the neutrino luminosity for an AGN SN Ia after the shock breakout is significantly larger than that of a classical SN IIn so that the difference for the amounts of their neutrino fluences may not be too significant.

It is expected that AGN stellar explosions could be ideal targets for future joint neutrino and EM multi-messenger  observations. We show the EM emission of the ejecta due to its interaction with the AGN disk materials and the SN Ia emission powered by radioactive decay, respectively, in Figure \ref{fig:MultiMessenger}, as predicted by \cite{grishin2021} and \cite{zhu2021c}. The neutrino burst may occur shortly after the shock breakout, while the associated EM signals last in much longer timescale, even up to several hundred days. Taking advantage of instantaneous EM follow-up observations of high-energy neutrino bursts can be conducive to search for these yet-discovered sources, which may provide smoking-gun evidence for the presence of stellar explosions embedded in AGN disks.

\section{Neutrino Fluence, {Detectability} and Diffuse Neutrino Emission}

{In Section \ref{sec:NeutrinoFluence}, we explore the impact of different explosion environments and different kinds of AGN stellar explosions, including SN Ia, core collapse SN (CCSN), GRB-SN, and kilonova, which are predicted to potentially occur embedded in AGN disks, on the neutrino fluence and individual detectablity. The results of diffuse neutrino emission and detection rates by IceCube for these AGN stellar explosions and choked GRBs with known theoretical event rate densities are presented in Section \ref{sec:DiffuseNeutrino}. }

\subsection{Neutrino Fluence and Individual Detectablity} \label{sec:NeutrinoFluence}
\begin{figure}
    \centering
    \includegraphics[width = 0.99\linewidth , trim = 65 30 95 60, clip]{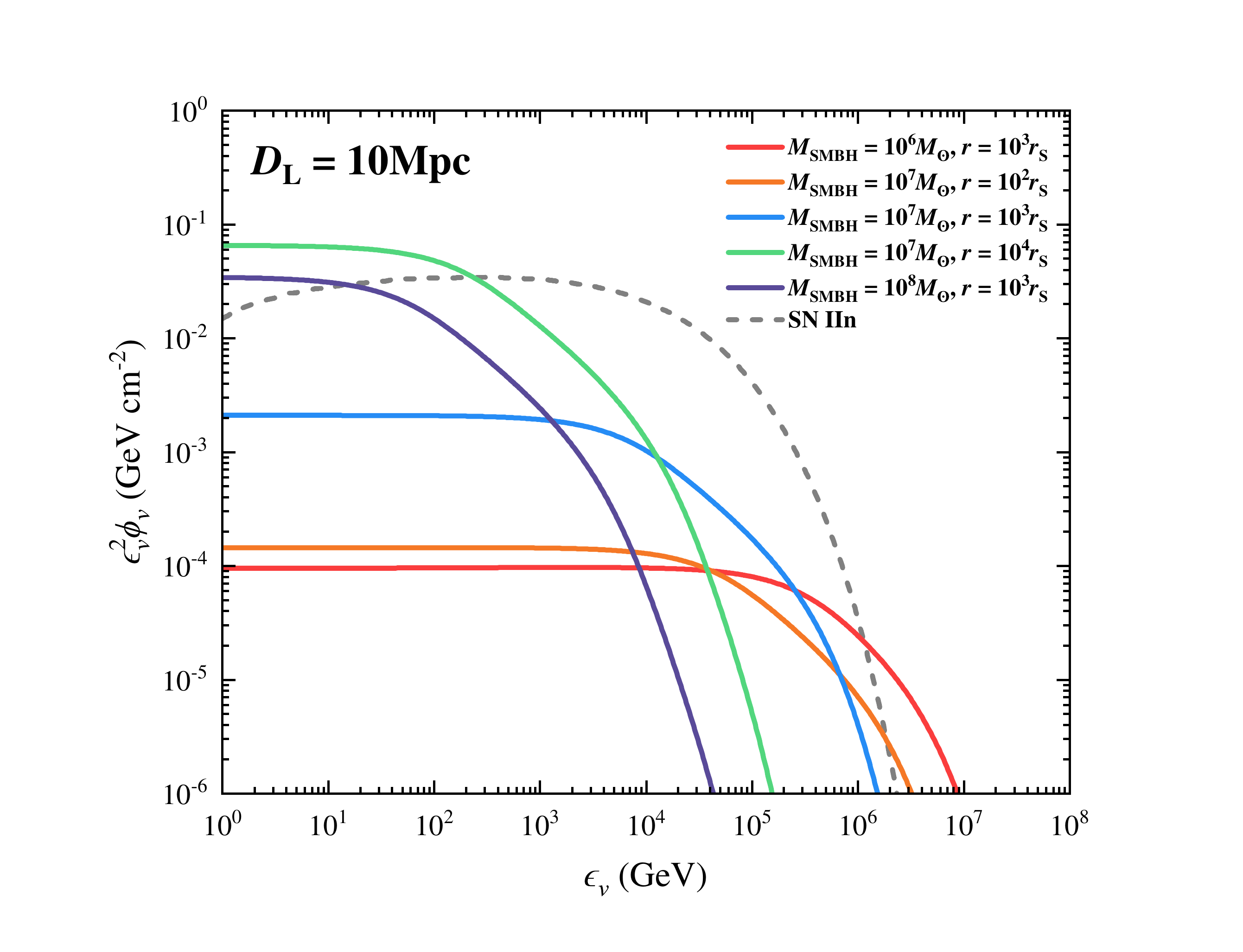}
    \caption{Energy fluences of all-flavor neutrinos from a single AGN SN Ia event occurring at $D_{\rm L} = 10\,{\rm Mpc}$. The colored lines (see labels for their meanings) represent different models with varied SMBH masses and radial locations. {For comparison,} the gray dashed line shows the neutrino spectra for a classical SN IIn from \cite{murase2018}.  }
    \label{fig:FluenceSMBHDependent}
\end{figure}

\begin{figure*}
   \centering
    \includegraphics[width = 0.49\linewidth , trim = 65 30 95 60, clip]{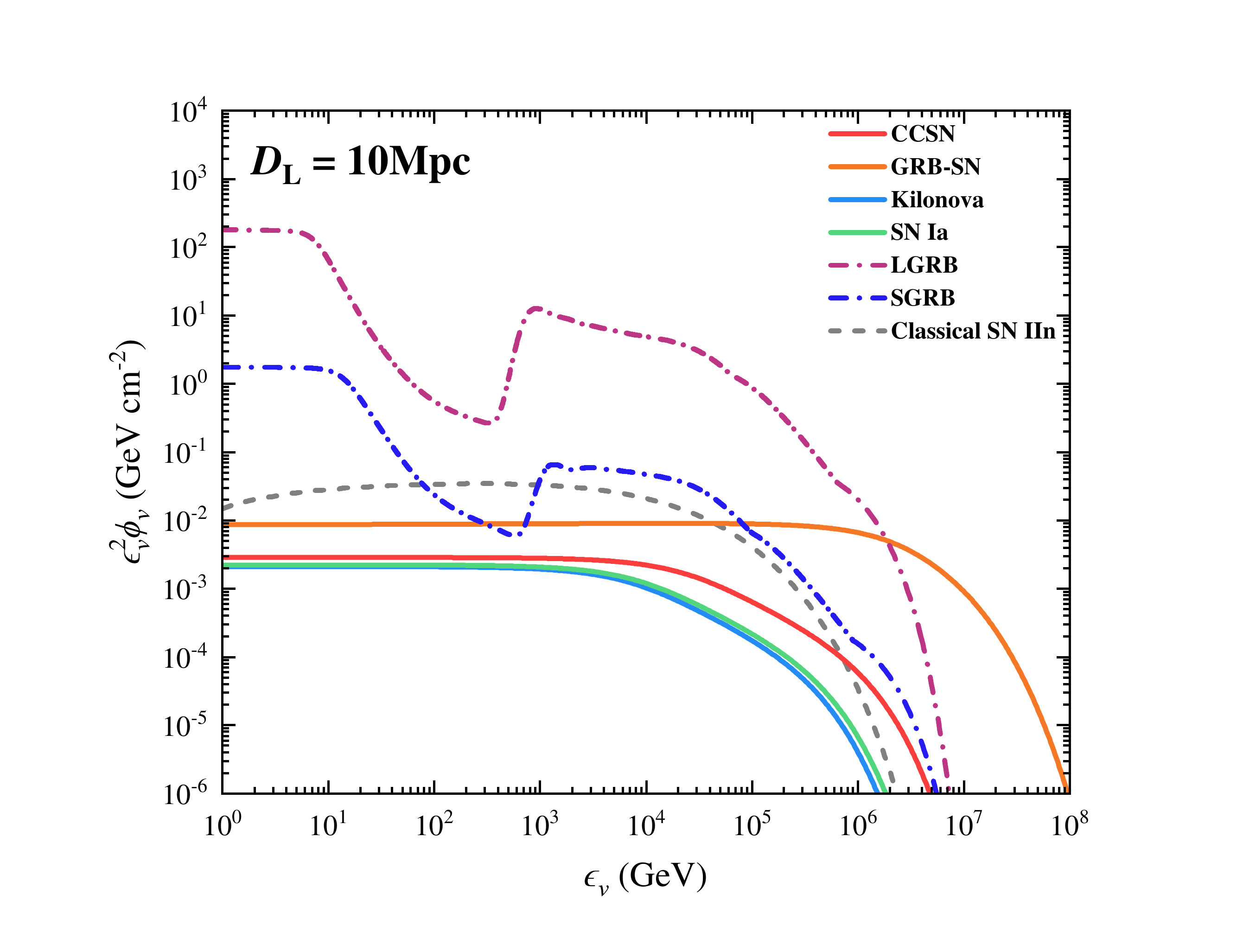}
    \includegraphics[width = 0.49\linewidth , trim = 65 30 95 60, clip]{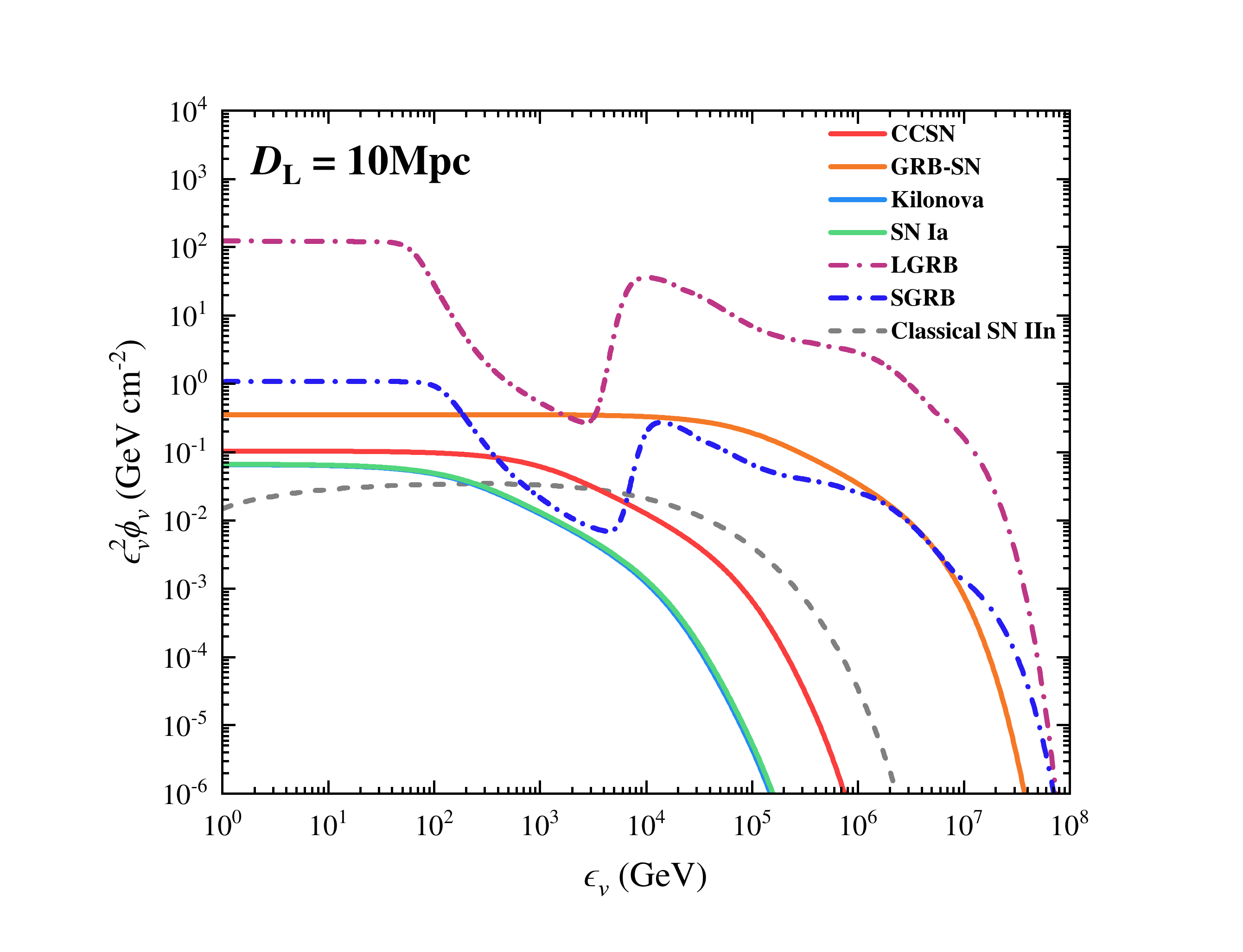}
    \caption{Energy fluences of all-flavor neutrinos from different AGN stellar explosions occurring at $r=10^3\,r_{\rm S}$ {(left panel)} and {$r=10^4\,r_{\rm S}$ (right panel)} around an SMBH of mass $M_\bullet = 10^7\,M_\odot$. The colored lines (see labels for meanings) represent different models with various kinds of AGN stellar explosions. For comparison, energy fluences from choked LGRBs and SGRBs are cacluated based on \cite{zhu2021b}. The gray dashed line shows the neutrino spectra for a classical SN IIn from \cite{murase2018}. The luminosity distance we assumed is $D_{\rm L} = 10\,{\rm Mpc}$. }
    \label{fig:FluenceEventDependent}
\end{figure*}

\begin{deluxetable*}{ccc|cccc|ccccccc|c}[htpb!] \label{Tab:1}
\tablecaption{Parameters for AGN stellar explosions}
\tablecolumns{14}
\tablewidth{0pt}
\tablehead{
\colhead{Explosion} &
\colhead{$E_{0,51}$\tablenotemark{a}} &
\colhead{$M_{\rm ej}$\tablenotemark{a}} &
\colhead{$M_{\bullet}$} &
\colhead{$r/r_{\rm S}$} &
\colhead{$\rho_{0,-9}$} & 
\colhead{$H_{12}$} &
\colhead{$\rho_{{\rm bo},-11}$} &
\colhead{$h_{\rm bo,14}$} &
\colhead{$v_{\rm bo} / c$} &
\colhead{$L_{\rm s,bo,45}$} &
\colhead{$E_{\rm s,48}$} &
\colhead{$\epsilon_{p,{\rm max}}^{\rm bo}/{\rm TeV}$} &
\colhead{$N_{\nu_\mu}$} &
\colhead{Remarks}
}
\startdata
SN Ia & $1$ & $1.3$ & $10^7$ & $10^3$ & $120$ & $23$ & $1.5$ & $1.0$ & $0.04$ & $1.1$ & $10$ & $91$ & $0.006$ & a,b \\
\hline
SN Ia  & $1$ & $1.3$ & $10^6$ & $10^3$ & $12000$ & $2.3$ & $3.93$ & $0.12$ & $0.16$ & $4.1$ & $0.64$ & $4100$ & $0.0007$ & a \\
SN Ia  & $1$ & $1.3$ & $10^7$ & $10^2$  & $36$ & $4.9$ & $3.0$ & $0.20$ & $0.08$ & $0.86$ & $0.82$ & $620$ & $0.0007$ & a \\
SN Ia  & $1$ & $1.3$ & $10^7$ & $10^4$ & $0.12$ & $350$ & $0.34$ & $9.4$ & $0.007$  & $0.21$ & $210$ & $2.4$ & $0.01$ & a \\
SN Ia  & $1$ & $1.3$ & $10^8$ & $10^3$ & $1.2$ & $230$ & $0.76$ & $8.7$ & $0.006$ & $1.3$ & $97$ & $0.74$ & $0.001$ & a \\
\hline
Kilonova & $1$ & $0.05$ & $10^7$ & $10^3$ & $120$ & $23$ & $1.5$ & $1.0$ & $0.04$ & $1.3$ & $11$ & $110$ & $0.006$ & b \\
Kilonova & $1$ & $0.05$ & $10^7$ & $10^4$ & $0.12$ & $350$ & $0.34$ & $9.4$ & $0.008$ & $0.21$ & $220$ & $2.5$ & $0.01$ & b \\
CCSN  & $3$ & $10$  & $10^7$ & $10^3$ & $120$ & $23$ & $1.0$ & $1.1$ & $0.05$ & $2.5$ & $15$ & $300$ & $0.01$ & b \\
CCSN  & $3$ & $10$  & $10^7$ & $10^4$ & $0.12$ & $350$ & $0.21$ & $11$ & $0.01$ & $0.70$ & $405$ & $13$ & $0.08$ & b\\
GRB-SN & $30$ & $10$ & $10^7$ & $10^3$ & $120$ & $23$ & $0.29$ & $1.1$ & $0.20$ & $51$ & $69$ & $26000$ & $0.07$ & b \\
GRB-SN & $30$ & $10$ & $10^7$ & $10^4$ & $0.12$ & $350$ & $0.06$ & $12$ & $0.05$ & $16$ & $2000$ & $1100$ & $2.1$ & b\\
\enddata
\tablecomments{The columns are [1] the kind of AGN stellar explosion; [2] explosion energy (in $10^{51}\,{\rm erg}\,{\rm s}^{-1}$); [3] ejecta mass (in $M_\odot$); [4] SMBH mass (in $M_\odot$); [5] radial location; [6] mid-plane radial density (in $10^{-9}\,{\rm g}\,{\rm cm}^{-3}$); [7] disk scale height (in $10^{12}\,{\rm cm}$); [8] disk density at the shock breakout (in $10^{-11}\,{\rm g}\,{\rm cm}^{-3}$); [9] breakout vertical height (in $10^{14}\,{\rm cm}$); [10] break out velocity; [11] the kinetic luminosity of the shock when the shock break out (in $10^{45}\,{\rm erg}\,{\rm s}^{-1}$); [12] shock energy after the shock breakout, i.e., $E_{\rm s} = \int_{t_{\rm bo}}^{+\infty}L_{\rm s}dt$ (in $10^{48}\,{\rm erg}$); [13] the maximum proton energy when the shock break out (in $1\,{\rm TeV}$); {[14] the number of up-going detected muon neutrinos in the IceCube from an single event at $D_{\rm L} = 10\,{\rm Mpc}$;} [15] remarks describing the parameters being varied: a) model with varied the SMBH mass and radial location; b) model with varied the kind of AGN stellar explosion.}
\tablenotemark{a}{References for the clasical explosion energy and ejecta mass of each stellar explosion: [1] SN Ia \citep{maoz2014}; [2] CCSN \citep{branch2017}; [3] GRB-SN \citep{cano2017}; [4] kilonova  \citep[based on the observations of GW170817/AT2017gfo, e.g.,][]{cowperthwaite2017,kasen2017,kasliwal2017,villar2017} }
\end{deluxetable*}

The neutrino fluence for a single event can be expressed as
\begin{equation}
    \epsilon_\nu^2\phi_\nu \approx \frac{1}{4\pi D_{\rm L}^2}\int_{t_{\rm bo}}^{+\infty}{\epsilon_\nu L_\nu} dt,
\end{equation}
where $dt=dh/v_{\rm s}$ and $D_{\rm L}$ is the luminosity distance. Figure \ref{fig:FluenceSMBHDependent} shows the all-flavor fluences of a single AGN SN Ia with varied SMBH masses and radial locations at $D_{\rm L} = 10\,{\rm Mpc}$. The grid of initial conditions and final shock breakout parameters are listed in Table \ref{Tab:1}. {We also estimate the number of muon neutrino events in the IceCube detector by}
\begin{equation}
    N_{\nu_\mu}(>1\,{\rm TeV}) = \int_{1\,{\rm TeV}}^
    {\epsilon_{\nu,{\rm max}}}d\epsilon_{\nu_\mu} A_{\rm eff}(\epsilon_{\nu_\mu})\phi_{\nu_\mu},
\end{equation}
{where $A_{\rm eff}(\epsilon_{\nu_\mu})$ is the effective area given by \cite{aartsen2017}. The number of up-going detected muon neutrinos from an single event located at $10\,{\rm Mpc}$ are shown in Table \ref{Tab:1}}.

From Figure \ref{fig:FluenceSMBHDependent}, we see that the neutrino fluence is roughly proportional to the mass of the SMBH and the radial location, which is also consistent to the total shock energy after the shock breakout, i.e., $E_{\rm s} = \int_{t_{\rm bo}}^{+\infty}L_{\rm s}dt$ shown in Table \ref{Tab:1}. Since the kinetic luminosity of the shock when it breaks out has a similar order of magnitude for different initial conditions, the shock driven by an event occurring at a larger radial location around a more massive SMBH can have a lower velocity and experience a slower surrounding disk density change. It thus can result in a slower evolution of the shock kinetic luminosity. More energy would thus be carried by the shock so that neutrino production would be more efficient, which can be more easy to produce a similar amount of neutrino fluence compared with classical SNe IIn. Furthermore, a lower velocity shock would lead to a lower maximum energy of protons and, hence, a lower maximum energy of neutrinos. AGN stellar explosions located at the outer parts of accretion disks around more massive SMBHs could be more difficult to produce higher-energy neutrinos, which is obviously shown in Figure \ref{fig:FluenceSMBHDependent}. {The number of detected neutrinos from an up-going single event with different SMBH masses and radial locations shown in Table \ref{Tab:1} is consistent with the neutrino fluence. }

We explore the impact of different kinds of AGN stellar explosions, including SN Ia, CCSN, GRB-SN, and kilonova, with the consideration of their classical explosion energies and ejecta masses which are listed in Table \ref{Tab:1}. {SNe Ia and kilonovae are more likely to occur near the trapping orbit, i.e., at $r\lesssim10^3\,r_{\rm S}$ \citep[e.g.,][]{bellovary2016,peng2021}. A large fraction of AGN stars could be formed {\em in situ} in the self-gravitating region, i.e., $10^3\lesssim r/r_{\rm S}\lesssim 10^5$ \citep[e.g.,][]{sirko2003,thompson2005}. Many of these stars would not be able to migrate to the inner trapped orbits of the disks before their deaths or within the AGN lifetime, so that AGN CCSNe and GRB-SNe could occur at the self-gravitating region of the disks. SMBHs with a mass of $\sim10^6-10^8\,M_\odot$ may be more common, which have a nearly uniform local mass function distribution \citep[e.g.,][]{kelly2012}. We may simply set all the events to occur around $10^7\,M_\odot$ SMBHs and consider two groups of potential radial locations, i.e., $r = 10^3\,r_{\rm S}$ and $10^4\,r_{\rm S}$.} As shown in Figure \ref{fig:FluenceEventDependent} {and Table \ref{Tab:1}, one can see that the same location} events with larger explosion energies can drive more powerful shocks and produce higher neutrino fluences with higher maximum neutrino energies. {At $D_{\rm L} = 10\,{\rm Mpc}$, only GRB-SNe occurring at larger radial locations of AGN disks could be easily detected by IceCube. The number of detected muon neutrinos for other AGN stellar explosions would be negligible. Unless these explosions have very high local event rate densities, they may be difficult to  discover by IceCube in the future.}

Relativistic jets from long or short GRB events in AGN discs can successfully break out from the stellar envelope and the kilonova ejecta, but would be choked during the propagation in the disk atmosphere. Such choked jets can also produce high-energy neutrinos as suggested by \cite{zhu2021b}. For comparison, we show that the neutrino fluences produced by a long-duration GRB (LGRB) jet with an isotropic energy of $E_{\rm iso} = 10^{53}\,{\rm erg}$ and a short-duration GRB jet (SGRB) with $E_{\rm iso} = 10^{51}\,{\rm erg}$ in Figure \ref{fig:FluenceEventDependent}. Different from the ejecta-disk interaction case studied here where $pp$ reaction is the main neutrino production process, for choked jets the neutrinos spectrum above $\sim1\,{\rm TeV}$ that we are interested in is mainly attributed to $p\gamma$ reactions. The dip around a few TeV is caused by the suppression of neutrino production due to the Bethe-Heitler process. Neutrino production from choked jets is much more efficient than that from ejecta-disk interactions, but their maximum neutrino energies of the two types of systems could be similar.

\subsection{Diffuse Neutrino Emission and Detection Rates}
\label{sec:DiffuseNeutrino}

\begin{deluxetable}{cccc}[htpb!] \label{Tab:2}
\tablecaption{Theoretical Event Rate Densities, {Diffuse Fractional Fluxes, and Detection Rates} for AGN Stellar Explosions}
\tablecolumns{3}
\tablewidth{0pt}
\tablehead{
\colhead{Explosion} &
\colhead{$R_0/{\rm Gpc}^{-3}\,{\rm yr}^{-1}$} &
\colhead{Diffuse Fractional Flux} &
\colhead{$\dot{N}/{\rm yr}^{-1}$}
}
\startdata
SN Ia & $<5000$ & $\lesssim3\%$ & $\lesssim0.1$\\
Kilonova & $<460$ & $\lesssim0.3\%$ & $\lesssim0.01$\\
SGRB & $<460$ & $\lesssim2\%$ & $\lesssim0.05$\\
CCSN & $<100$ & $\lesssim1\%$ & $\lesssim0.03$ \\
GRB-SN & $<1$ & $\lesssim0.3\%$ & $\lesssim0.01$\\
LGRB & $<1$ & $\lesssim5\%$ & $\lesssim0.2$
\enddata
\tablecomments{We assume all binary WD mergers can produce SNe Ia, while all binary NS mergers and $\sim20\%$ NS--BH mergers can power SGRBs and kilonovae \citep{mckernan2020}. {The local event rate density for AGN CCSN is based on the constraint by \cite{grishin2021}. The rate densities of AGN GRB-SNe and LGRBs are assumed to be $\sim1\%$ of AGN CCSNe since \cite{jermyn2021} and \cite{dittmann2021} suggested that AGN stars could have extremely high spins and easily make LGRBs embedded in AGN disks. } }
\end{deluxetable}

\begin{figure}
    \centering
    \includegraphics[width = 0.99\linewidth , trim = 65 30 95 60, clip]{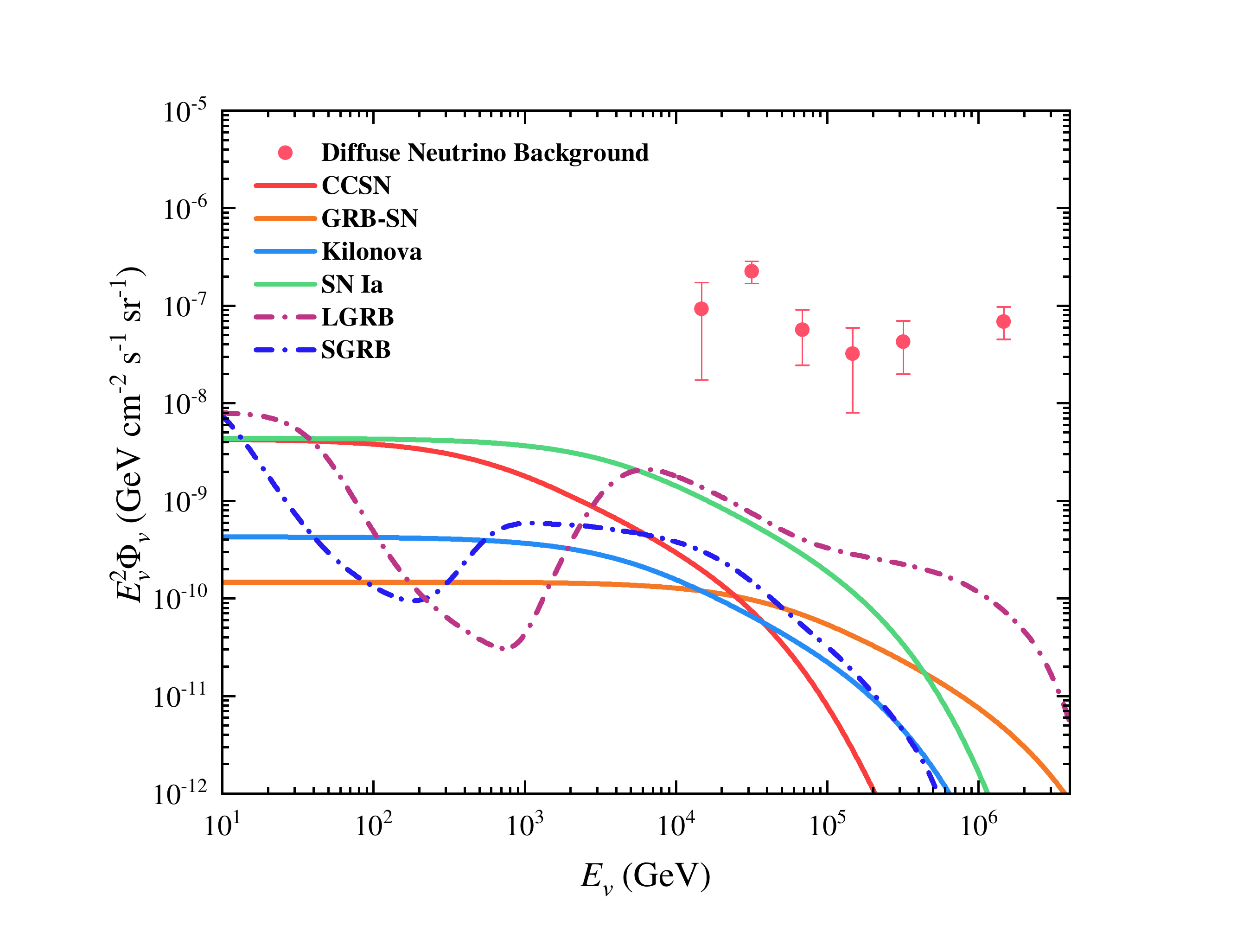}
    \caption{The colored lines (see labels for meanings) represent the upper limits of expected all-flavor diffuse neutrino fluences contributed from {AGN stellar explosions}. The pink circles are observed astrophysical diffuse neutrino fluence measured by IceCube \citep{aartsen2015}. }
    \label{fig:FluenceDiffusedSpectrum}
\end{figure}

The diffuse neutrino fluence can be calculated by \citep[e.g.,][]{razzaque2004,xiao2016}
\begin{equation}
    E_\nu^2\Phi_\nu = \frac{c}{4\pi H_0}\int_0^{z_{\rm max}}\frac{R_0f(z)\epsilon_\nu^2\phi_\nu(\epsilon_\nu)}{(1 + z)^2\sqrt{\Omega_{\rm m}(1 + z)^3 + \Omega_\Lambda}}dz,
\end{equation}
where $z$ is the redshift, $E_\nu = \epsilon_\nu / (1 + z)$ is the neutrino energy in the observer's frame, $R_0$ is the local event rate density, and $f(z)$ is the redshift distribution. The standard $\Lambda$CDM cosmology with $H_0 = 67.8\,{\rm km}\,{\rm s}^{-1}\,{\rm Mpc}^{-1}$, $\Omega_{\rm m} = 0.308$, and $\Omega_\Lambda = 0.692$ \citep{planck2016} is applied. In order to estimate the diffuse neutrino fluence from AGN stellar explosions, one needs to know the local event rate density and redshift distribution for each kind of event. At present, the event rate densities for some kinds of AGN stellar explosions are predicted (see Table \ref{Tab:2}). {Following the discussion in Section \ref{sec:NeutrinoFluence}, we may simply set all SNe Ia, kilonovae, and SGRBs to occur at $r = 10^3\,r_{\rm S}$ around $10^7\,M_\odot$ SMBHs while all CCSNe, GRB-SNe, and LGRBs to occur at $r = 10^4\,r_{\rm S}$ around $10^7\,M_\odot$ SMBHs.} Because the cosmic evolution of AGN and star formation rate is not significant \citep[e.g.,][]{madau2014}, we roughly assume that stellar explosions could closely track the star formation history, which may be expressed by the redshift-evolution factor \citep[e.g.][]{sun15} $f(z) = \left[(1+z)^{3.4\eta} + \left(\frac{1+z}{5000}\right)^{-0.3\eta} + \left(\frac{1+z}{9}\right)^{-3.5\eta} \right]^{1/\eta}$, where $\eta = -10$ \citep{yuksel2008}. The predicted diffuse neutrino fluences by considering these AGN stellar explosions are shown in Figure \ref{fig:FluenceDiffusedSpectrum}. {For each AGN stellar explosion, we simulate $1\times10^6$ events in the universe to estimate the detection rate. The corresponding simulated detection rates $\dot{N}$ for different AGN stellar explosions are listed in Table \ref{Tab:2}. } These kinds of stellar explosions may contribute to $\lesssim10\%$ of the observed neutrino background.

\section{Discussion}

In this {\em Letter}, we show that high-energy neutrinos can be produced during interactions between the ejecta from AGN stellar explosions and AGN disk materials. The neutrino signal has a shorter duration compared with the EM signal, due to the rapid drop of the mass density above the disk. {Two processes may increase the duration of the neutrino signals. On one hand,} it is likely that there exist disk winds \citep[e.g.,][]{proga2000,proga2004} and/or broad-line region (BLR) clouds \citep[e.g.,][]{moriya2017} surrounding the accretion disks. The ejecta shock, after interacting with disk materials, may continue to propagate in the disk winds and/or BLR clouds, typically having similar densities to those found in SNe IIn. The rate of neutrino production after shock breakout would decay more slowly compared with the presented predictions. {On the other hand, cavities or even open gaps \citep[e.g.,][]{kimura2021,wang2021a,wang2021b} are predicted to exist around the orbits of AGN objects. The densities of cavities and gaps are much lower than the density of disk atmosphere, which would reduce the swept mass when the shock breaks out for AGN stellar explosions. Therefore, the final expected breakout velocities could be much higher than our predictions. In these two cases, the expected neutrino fluence of a single AGN stellar explosion and total diffuse neutrino contributions would be higher than predicted.}

For the calculations of the diffuse neutrino background emission, we show that SNe Ia, kilonovae, CCSNe, GRB-SNe, and choked GRBs in AGN disks may contribute to $\lesssim10\%$ of the observed diffuse neutrino background. SNe \citep{grishin2021}, Bondi explosions of stellar BHs \citep{wang2021a} and mTDEs \citep{yang2021} are predicted to have very high event rate densities in AGN disks. Such energetic events can contribute more considerably to the diffuse neutrino background. Despite of the uncertainties, one may still expect that AGN stellar explosions could potentially make an important contribution to the astrophysical neutrino background. Future more detailed theoretical studies and observable constraints for the event rate densities of AGN stellar explosions can give a better estimation of their contribution to the neutrino background.

\acknowledgments

We thank Zhuo Li, B. Theodore Zhang, Daniel Proga, Jian-Min Wang, He Gao, Yun-Wei Yu, and Ming-Yang Zhuang for valuable comments. The work of J.P.Z is partially supported by the National Science Foundation of China under Grant No. 11721303 and the National Basic Research Program of China under grant No. 2014CB845800. K.W is supported by the National Natural Science Foundation under grants 12003007 and the Fundamental Research Funds for the Central Universities (No. 2020kfyXJJS039).

\bibliography{neutrino}{}
\bibliographystyle{aasjournal}

\end{document}